\documentclass [twocolumn, aps, showpacs] {revtex4}
\usepackage{graphicx}
\begin{document}
\draft
\title {Layer charge instability in unbalanced bilayer systems in the quantum Hall regime}

\author {E. Tutuc, R. Pillarisetty, S. Melinte, E.P. De Poortere, and M. Shayegan}
\address{Department of Electrical Engineering, Princeton
University, Princeton, NJ 08544}
\date{\today}

\begin{abstract}
Measurements in GaAs hole bilayers with unequal layer densities
reveal a pronounced magneto-resistance hysteresis at the magnetic
field positions where either the majority or minority layer is at
Landau level filling factor one. At a fixed field in the
hysteretic regions, the resistance exhibits an unusual time
dependence, consisting of random, bidirectional jumps followed by
slow relaxations. These anomalies are apparently caused by
instabilities in the charge distribution of the two layers.
\end{abstract}
\maketitle

Hysteretic phenomena are widespread in nature. They are common
magnetic materials, and often indicate a non-equilibrium situation
associated with a phase transition and the presence of domains
\cite{berttoti}. Recently, hysteresis has also been reported in
various two-dimensional (2D) carrier systems in semiconductor
structures at low temperatures and high magnetic fields
\cite{cho,kronmueller,piazza,eom,etienne,smet}. In these cases,
magneto-resistance ($\rho_{xx}$) hysteresis appears in the quantum
Hall (QH) regime when two Landau levels (LLs) with opposite spin
are brought into coincidence. While the 2D systems studied have
been notably different, the common thread in these experiments is
that there is a {\it magnetic} transition involving the carrier
{\it spin} \cite{tomas}.

Here we present hysteretic $\rho_{xx}$ data in 2D bilayer systems
in the QH regime. The hysteresis in these systems has a different
origin and is caused by a non-equilibrium charge distribution in
the two layers. We studied the magneto-transport coefficients of
GaAs bilayer hole systems with unequal layer densities.  When the
interlayer tunneling is sufficiently small, $\rho_{xx}$ of the
bilayer system exhibits a pronounced hysteresis at perpendicular
magnetic field ($B$) positions close to where either the majority
or minority layer is at LL filling factor one. Most remarkable is
the time dependence of $\rho_{xx}$ at a fixed field in the
hysteretic regime, when the two layers are closely spaced. As a
function of time, $\rho_{xx}$ exhibits large, random, sudden jumps
toward higher and lower values, followed by a slow decay in the
opposite direction. The data may signal an instability in the
charge distribution of the two layers, i.e., an instability
associated with the {\it pseudospin} (layer), rather than spin,
degree of freedom.

We studied nine GaAs bilayer hole samples from six different
wafers, all grown on GaAs (311)A substrates and modulation doped
with Si. In all samples, the holes are confined to two 15nm-wide
GaAs quantum wells which are separated by AlAs or AlAs/AlGaAs
barriers with thickness $7.5\leq W \leq 200$nm. The rather thick
barrier combined with the large effective mass of GaAs 2D holes
\cite{mass} reduces considerably the tunneling between the two
layers \cite{tunneling}. As grown, the samples have layer
densities of $\leq7\times10^{10}$ cm$^{-2}$, and low temperature
($T$) mobilities of $\sim35$ m$^2$/Vs. Metallic top and bottom
gates were added to control the densities in the layers. We
studied several types of devices, including 2.5{$\times$}2.5mm
square samples and ones with patterned Hall bars; in these samples
the ohmic contacts contact both layers. One sample was fabricated
using a selective depletion scheme \cite{eisenstein-APL} that
allows probing the transport characteristics of individual layers.
The measurements were performed in a dilution refrigerator down to
$T=20$mK.

\begin{figure*}
\centering
\includegraphics[scale=0.58]{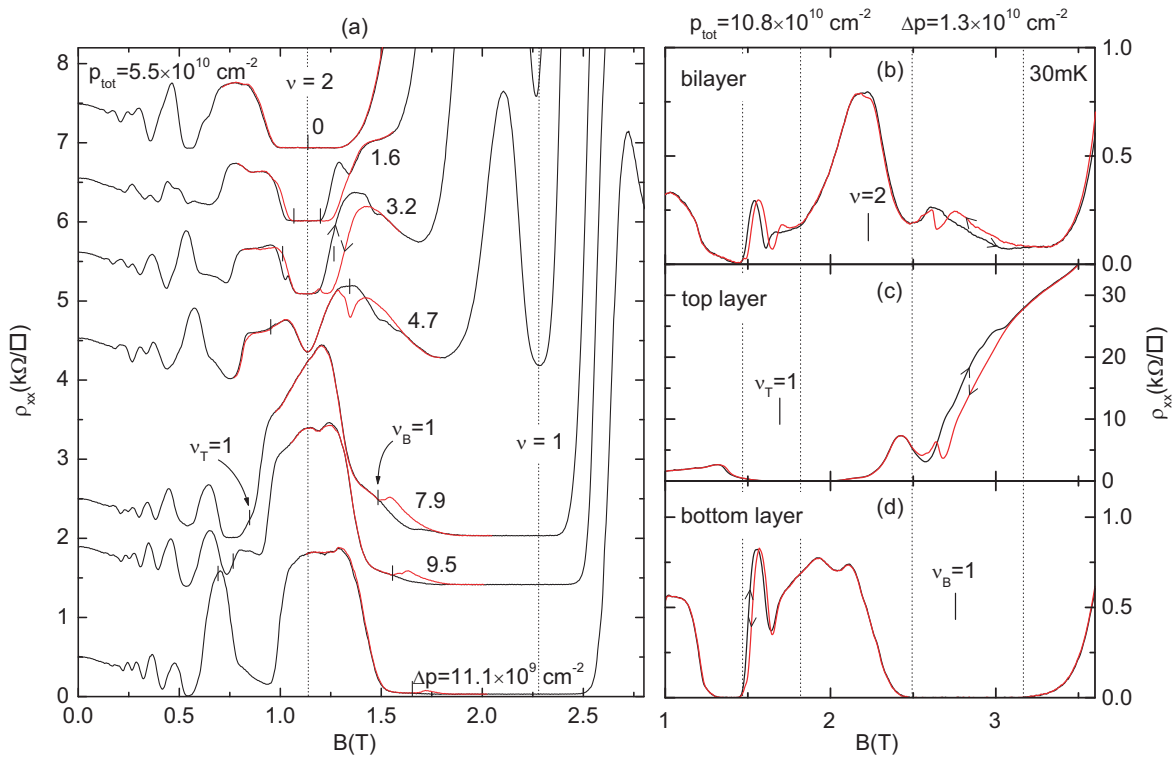}
\caption{\small{(a) $\rho_{xx}$ vs. $B$ traces for a GaAs bilayer
hole sample with a barrier width $W=11$nm. Data are shown for
different values of charge transfer, $\Delta p$, while $p_{tot}$
is kept constant at $5.5\times10^{10}$ cm$^{-2}$. The traces are
shifted vertically for clarity. The dotted lines indicate the
positions of bilayer filling factors $\nu=2$ and $\nu=1$. A strong
hysteresis develops in $\rho_{xx}$ when the bilayer system is
imbalanced. The right and left tick marks in each trace indicate
the estimated positions of filling factor one for layers with
higher (bottom) and lower (top) densities, respectively. The right
panel shows data for a similar sample but with independent layer
contacts, and $p_{tot}=10.8\times10^{10}$ cm$^{-2}$ and $\Delta
p=1.3\times10^{10}$ cm$^{-2}$. (b) $\rho_{xx}$ vs. $B$ traces
measured when the ohmic contacts are connected to both layers. (c)
and (d) $\rho_{xx}$ of the top and bottom layers measured
individually. In all figures the black (red) line represents the
trace taken when $B$ is swept up (down).}}
\end{figure*}

Data of Fig. 1 highlight some of the results of our study. In (a)
we show a set of traces where $\rho_{xx}$ was measured for a
sample with $W=11$nm as $B$ was ramped up or down. For the traces
of Fig. 1(a) the total bilayer density ($p_{tot}$) is kept
constant at $5.5\times10^{10}$ cm$^{-2}$ while charge is
transferred from one layer to another using back- and front-gate
biases. We define the charge transfer from one layer to another as
$\Delta p=(p_{B}-p_{T})/2$, where $p_{B}$ and $p_{T}$ are the
densities of bottom and top layers, respectively \cite{note2}. At
a given value of $B$, we define the filling factor, $\nu$, of the
bilayer system as the ratio between $p_{tot}$ and the LL
degeneracy, $eB/h$. We also introduce the filling factors for top
and bottom layers, $\nu_{T}$ and $\nu_{B}$ respectively, as the
ratio between the layer density and $eB/h$.

The data of Fig. 1(a) show that when the bilayer system is
balanced ($\Delta p$=0; top trace) $\rho_{xx}$ is independent of
the direction $B$ is ramped. However, as soon as the system is
imbalanced ($|\Delta p|>0$) a strong hysteresis develops in
$\rho_{xx}$. For values of $\Delta p<4.7\times10^{9}$ cm$^{-2}$,
$\rho_{xx}$ displays hysteretic behavior in two field ranges near
$\nu=2$, one near  $\nu_{T}=1$ and another near $\nu_{B}=1$. When
$\Delta p\geq4.7\times10^{9}$ cm$^{-2}$, the hysteresis exists
only near $\nu_{B}=1$. For sufficiently large $\Delta p$, no
hysteresis is observed. The amplitude of the hysteresis also
decreases as $T$ is increased (data not shown) and vanishes
completely above $T\simeq$230mK, roughly independent of $\Delta
p$.

To probe the contribution of the spin degree of freedom to the
hysteresis in our bilayer systems, we performed measurements in
tilted magnetic fields on a sample very similar to the one shown
in Fig. 1(a). In this experiment the direction of the field was
kept at an angle $\theta$ with respect to the normal to the plane
of the bilayer system. For $\theta$ ranging from $0^{\circ}$ to
$80^{\circ}$, corresponding to a six-fold increase of the total
field (and therefore of the Zeeman energy) in the hysteretic
region, the position of the hysteresis in perpendicular magnetic
field did not change at all. If the hysteresis were caused by an
instability associated with the spin degree of freedom, one would
expect \cite{etienne, smet} that the applied parallel field would
change the position and magnitude of the hysteresis. These results
rule out spin as being responsible for the hysteresis.

To better understand the origin of the observed hysteresis, we
fabricated another sample from a different wafer, also with
$W=11$nm, using a selective depletion scheme
\cite{eisenstein-APL}, and aimed to independently probe each layer
of the bilayer system. The data are shown in the right panel of
Fig. 1. In (b) we plot $\rho_{xx}$ for the bilayer system, that
is, when the ohmic contacts are connected to both layers, for both
up and down $B$-sweeps. In (c) and (d) we show $\rho_{xx}$ traces
for the top and bottom layers, measured separately, but at the
same pair of layer densities as in panel (b) traces. Two features
of these data are noteworthy. First, the traces of (b) exhibit
hysteresis in two ranges of $B$. The hysteresis between 1.45 and
1.8T matches well the position of $\nu_{T}=1$ QH state of the top
layer as seen in (c), while the hysteresis located between 2.5 and
3.2T overlaps the QH state of the bottom layer (see (d)). This
observation confirms that the hysteresis in $\rho_{xx}$ of the
bilayer system takes place when one of the layers is at filling
factor one. Second, data of (c) and (d) show that each individual
layer exhibits hysteresis when the other layer is at filling
factor one, i.e., $\rho_{xx}$ of the top layer exhibits hysteresis
when $\nu_{B}=1$, and vise versa.

\begin{figure}
\centering
\includegraphics[scale=0.42]{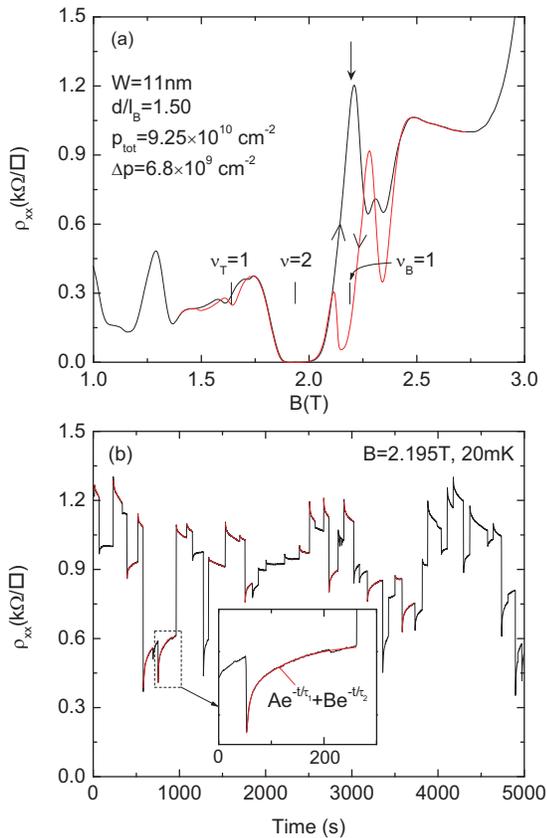}
\caption{\small{(a) $\rho_{xx}$ vs. $B$ traces for another sample
with $W=11$nm, with $p_{tot}$ and $\Delta p$ as indicated. (b)
$\rho_{xx}$ vs. time measured following an up-sweep of $B$ and
stopping $B$ at 2.195T. $\rho_{xx}$ displays sudden jumps followed
by a slow relaxation. The red lines represent double exponential
fits to the slow relaxation component of the data, with time
constants $\tau_{1}\sim5s$ and $\tau_{2}\sim90s$.}}
\end{figure}

The observation of hysteretic magneto-resistance in unbalanced
bilayer systems has precedence. Zhu {\it et al.} \cite{zhu}
reported hysteresis in 2D electron systems with a parallel
conducting layer. In their case, the parallel layer was a
parasitic, low-mobility, doping layer at a distance of 75 to 95 nm
away from the high-mobility 2D electrons.  Similar to our data,
they observed hysteresis in $\rho_{xx}$ when the layer containing
the high-mobility 2D electrons was in a QH state. They also
presented a simple model to explain the observed hysteresis. In
their model, the hysteresis comes about because of a
non-equilibrium charge distribution in the layers. As $B$ is
swept, thanks to the Landau quantization, the Fermi levels of both
layers oscillate. These oscillations lead to temporary imbalances
between the chemical potentials of the two layers. The potential
imbalance is particularly abrupt and large when the high-mobility
2D electrons enter a QH state as their Fermi level jumps by a
significant amount, equal to the separation between the adjacent
LLs. With increasing time, of course, the Fermi levels of the two
layers have to come to equilibrium since the latter are shorted
together via the ohmic contacts. But this equilibration can take a
long time in the QH effect regime: it has to take place via the
layers' edges and the ohmic contacts since the (bulk) states in
the center of the 2D layer that is in the QH state are localized
and the layer sheet conductivity is very small. As a result,
$\rho_{xx}$, which is recorded as $B$ is swept at a finite rate,
can show a hysteretic behavior. Consistent with their model, Zhu
{\it et al.} found that when the $B$-sweep is interrupted in the
hysteretic region and $\rho_{xx}$ is monitored as a function of
time, it decays approximately exponentially toward an equilibrium
value. Moreover, they found the time constant of the decay in
reasonable agreement with estimates based on the parameters of
their experiment.

\begin{figure}
\centering
\includegraphics[scale=0.61]{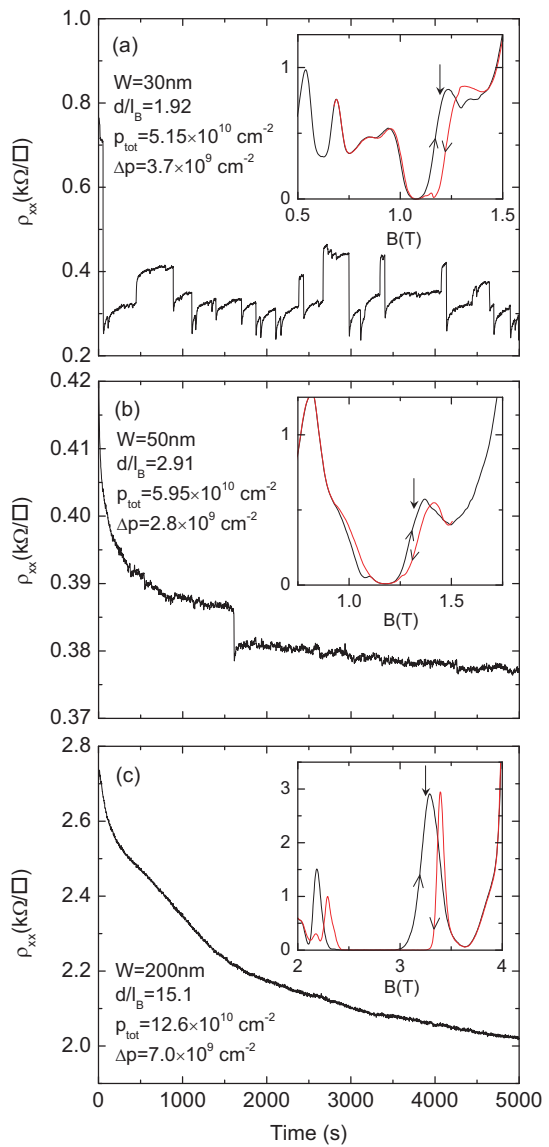}
\caption{\small{Time evolution of $\rho_{xx}$ in the hysteretic
region for samples with different $W$, $p_{tot}$, and $\Delta p$,
as indicated. The insets show $\rho_{xx}$ vs. $B$ for up- and
down-sweeps. The arrow in each inset indicates the position at
which the time evolution was recorded after an up-sweep of $B$.}}
\end{figure}

The hysteresis in our samples resembles what Zhu {\it et al.}
observe and likely has a similar origin. Our data of Fig. 1 in
fact explicitly show that the hysteresis happens when one of the
layers is in a QH state, and it is the resistivity of the layer
which is not in the QH state that is hysteretic. The
interpretation that the hysteresis indicates a charge transfer
between the two layers is also supported by our data: the
hysteresis observed in $\rho_{xx}$ of either the top or the bottom
layer measured separately appears much like a horizontal shift of
the trace in $B$, as if the layer had slightly different density
when ramping $B$ up or down [see, e.g., Fig. 1(c)]. The time
evolution of $\rho_{xx}$ in our samples, however, is qualitatively
different from the observations of Zhu {\it et al.}, and points to
a very unusual relaxation. In the remainder of the paper, we
describe this evolution and speculate on its possible origin.

We have studied a number of samples with varying $p_{tot}$,
$\Delta p$, and barrier width $7.5\leq W \leq 200$nm. Samples with
$W=7.5$nm do not show any hysteresis; this is likely because the
interlayer tunneling is sufficiently large so that the two layers
stay in equilibrium during the $B$-sweep. We observe hysteresis
for all samples with $11\leq W\leq200$nm, but the time evolution
of $\rho_{xx}$ critically depends on $W$. Examples are shown in
Figs. 2 and 3. For samples with $W=11$nm (Fig. 2), the time
evolution is simply wild! It displays sudden jumps in $\rho_{xx}$,
followed by a slow relaxation after each jump. Note that
$\rho_{xx}$ jumps toward {\it both} higher and lower values,
reminiscent of a bistability, although the jumps do not happen
between fixed values of $\rho_{xx}$. In between jumps, $\rho_{xx}$
follows a slow relaxation, in the opposite direction of the jump,
that can be fitted well by a double exponential [see Fig. 2(b)
inset]. It is noteworthy that even when measured over days (up to
2.5$\times10^5$s), we did not observe any tendency towards a
settling of the jumps. We wish to emphasize that, outside the
hysteretic region, $\rho_{xx}$ is independent of time to within
less than 0.3\%.

We have attempted to quantify the characteristics of this
evolution by two parameters: average frequency and amplitude of
the jumps. Our $T$-dependence measurements show that, at a fixed
$B$, the average jump amplitude decreases as $T$ increases. Not
surprisingly, the jumps are no longer visible above the
temperature where the hysteresis vanishes. On the other hand, the
average frequency appears to be independent of $T$. Also, the jump
frequency and amplitude are independent of the magnitude of the
sample current, as long as the current is kept sufficiently small
($\leq10$ nA) \cite{jouleheating}. Interestingly, the jumps and
decays appear to continue even when the current is completely
turned off \cite{note3}.

The sample with $W=30$nm shows a behavior qualitatively similar to
the one with $W=11$nm, although both the frequency and size of
$\rho_{xx}$ jumps are smaller [Fig. 3(a)]. Data for the $W=50$nm
sample, however, are qualitatively different [(Fig. 3(b)]: there
are by far fewer $\rho_{xx}$ jumps (typically one jump every few
$10^3$ s), and $\rho_{xx}$ appears to decay with time. Finally,
for the $W=200$nm sample [Fig. 3(c)], we typically observe a
simple decay with time \cite{jumps-in-200nm}. Interestingly, for
the samples of Figs. 3(b) and (c), $\rho_{xx}$ continues to decay
with an ever increasing time constant.

The time evolutions we observe for the sample with $W=200$nm
barrier [Figs. 3(c)] is qualitatively similar to the observation
of Zhu {\it et al.} \cite{zhu}. The time dependences for the
samples with barrier widths $W=11$ and 30nm, however, are very
unusual and cannot be understood in a simple model where the
bilayer system slowly and steadily relaxes to an equilibrium
state. In these samples, $\rho_{xx}$ displays sudden jumps that do
not have any tendency to settle, at least over a time scale of
days. We do not know the origin of these time evolutions. The
sudden jumps in $\rho_{xx}$ in our bilayers with small $W$ bear
some resemblance to the so-called Barkhausen jumps, which are
observed in magnetic materials \cite{berttoti}. The Barkhausen
jumps occur when the magnetic system finds a lower energy state
available and one or several domains change orientation. A
tantalizing speculation is that in the bilayer systems with close
layer separation, the interlayer interaction acts as an opposing
force to the charge transfer caused by the Fermi level difference
\cite{dlb} . In this scenario, the two opposing mechanisms may
mediate the creation of a complicated layer charge density pattern
or {\it pseudospin} domains.

Another possibility is that the observed jumps are not intrinsic
to the sample, but rather triggered by external sources (e.g.
electromagnetic noise). If so, it is a puzzle why the jumps are
much more frequent in the bilayer samples with smaller layer
separation and are seen only in the hysteretic region.

In summary we report an unusual time dependence associated with
hysteretic magneto-resistance in GaAs bilayer holes with close
layer separation. The resisitivity exhibits sudden jumps with
time, possibly caused by a layer charge instability.

We thank DOE and NSF for support and D. Haldane for helpful
discussions.


\begin{thebibliography}{10}
\bibitem{mass}
The carrier mass in GaAs 2D holes is $0.38m_{e}$, while that of
GaAs 2D electrons is $0.067m_{e}$; $m_{e}$ is the electron mass.
\bibitem{note2}
The total density is kept constant to within 3\%. The charge
transfer $\Delta p$ is measured with an accuracy of 6\%.
\bibitem{jouleheating}
At higher currents the average amplitude of the jumps gradually
decreases, likely because of Joule heating of the carriers.
\bibitem{jumps-in-200nm}
Even the $W=200$nm sample shows $\rho_{xx}$ jumps, but these are
extremely rare. We measured four traces like in Fig. 3(c), taken
over a total period of $5\times10^4$ s. In two traces we observed
a single jump, followed by a decay.
\bibitem{dlb}
As a measure of interalyer interaction, in Figs. 2 and 3, we have
included values for $d/l_B$, the ratio between the interalyer
distance ($d$) and the magnetic lenght ($l_B$).
\end{thebibliography}
\end{document}